# A Rapid Compression Expansion Machine (RCEM) for studying chemical kinetics: Experimental principle and first applications

M. Werler, R. Schiessl, U. Maas

ITT, Institute of Technical Thermodynamics, Karlsruhe Institute of Technology, Karlsruhe, Germany

Corresponding Author:

Marc Werler

Institute for Technical Thermodynamics

Karlsruhe Institute for Technology

76131 Karlsruhe, Germany

Phone: +49 (0) 721 608 42881

Email: marc.werler@kit.edu

Colloquium: Reaction Kinetics

**Word equivalent length of**

main text:            3359 words

equations:            137 words

references:           455 words

figures + captions:   fig. 1: 212 words

                                    fig. 2: 142 words

                                    fig. 3: 256 words

                                    fig. 4: 156 words

                                    fig. 5: 152 words

fig. 6: 159 words

fig. 7: 156 words

fig. 8: 217 words

fig. 9: 208 words

fig.10: 208 words

**Total Length of Paper: 5817 words** (used Method 1 as indicated in the Instructions to Authors for Manuscript Preparation)

# A Rapid Compression Expansion Machine (RCEM): Experimental principle and first applications

M. Werler, R. Schiessl, U. Maas

ITT, Institute of Technical Thermodynamics, Karlsruhe Institute of Technology, Karlsruhe, Germany

**Abstract**

A novel extension of a rapid compression machine (RCM), namely a Rapid Compression Expansion Machine (RCEM), is described and its use for studying chemical kinetics is demonstrated. Like conventional RCMs, the RCEM quickly compresses a fuel/air mixture by pushing a piston into a cylinder; the resulting high temperatures and pressures initiate chemical reactions. In addition, the machine can rapidly expand the compressed gas in a controlled way by pulling the piston outwards again. This freezes chemical activity after a pre-defined reaction duration, and therefore allows a convenient probe sampling and *ex-situ* gas analysis of stable species. The RCEM therefore is a promising instrument for studying chemical kinetics, including also partially reacted fuel/air mixtures. The setup of the RCEM, its experimental characteristics and its use for studying chemical reactions are outlined in detail. To allow comparisons of RCEM results with predictions of chemical reaction mechanisms, a simple numerical model of the RCEM process is described. As a first application, ex-situ measurements of the temporal evolution of species in partially reacted dimethyl ether/methane-mixtures are presented, and the results are compared to predictions of a reaction mechanism.

# 1. Introduction

Ignition delay time (IDT) is an important quantity for characterizing the auto-ignition behaviour of a fuel. A frequent use of IDT measurements is for the validation of reaction mechanisms [1-3]. While correct prediction of IDTs is a necessary requirement for a valid reaction mechanism, it is not sufficient. For a more extensive validation, additional quantities like flame speeds or species formation, have to be considered [2, 4-6]. Most measurements of temporal species profiles are conducted in flow reactors or shock tubes. In flow reactors, species time profiles can be measured by probe sampling using, for example, gas chromatography [8] or Fourier transform infrared spectrometry [9]. Davidson et al. measured time resolved multiple species by laser absorption in their shock tube [7]. Also: TOF-MS in shock tubes.

To access lower temperatures and longer reaction times for species measurements, rapid compression machines (RCMs) have been employed. Minetti et al. [10] reported species measurements during the ignition delay period of n-heptane. They rapidly quench the reacting mixture by bursting a diaphragm, allowing the gas to quickly expand into a collection vessel, and subsequently analyse the sample by gas chromatography and mass spectroscopy (GC/MS). Karwat et al. [11] compared species time profiles of n-heptane air mixtures in their rapid compression facility with predictions of a reaction mechanism. They used a fast valve to take probes of the test gas and analysed it by GC/MS, and demonstrated the benefit of an additional validation by means of temporal profiles of intermediate species and products [11].

In our study, an RCM-based device for stimulating and subsequent "quenching" of chemical reactions and measuring species profiles is introduced: A Rapid Compression Expansion Machine (RCEM), in which the piston can be moved both in- and outwards in the cylinder in a controlled way. This allows for quick compression of a gas (like an RCM), but also a subsequent quick, controlled expansion after a pre-selected time.

In the following, details of the RCEM setup will be given; to characterize the RCEM process, pressure histories from compression/expansion sequences, as well as their implications for interpreting RCEM measurements, are discussed. A simple computational model for describing the evolution of the cylinder load during an RCEM process is explained. Species measurements of partly reacted mixtures are then compared to predictions of a detailed reaction mechanism. As a first application of the RCEM technology, species measurements in methane/dimethyl ether mixtures under fuel rich conditions ($\phi$ = 2, 6 and 10) are shown and compared to model predictions using detailed chemistry. This highlights the capability of the RCEM as a device for investigating low-temperature chemistry.

The employed reaction mechanism is a combination of existing reaction mechanisms describing DME/$CH_4$ chemistry [12]. The reactions describing the chemistry of DME were taken from the mechanism of Zhao et al. [9]. Reactions of two mechanisms were combined to describe the range between usual combustion regimes [13] to pyrolysis of methane [14].

## 2. Experimental Setup

The RCEM test rig is based on a RCM, as explained in Werler et al. [15]. In the following, a brief description of the basic RCM-setup will be given. Modifications of the setup and the new expansion mechanism system are explained in more detail.

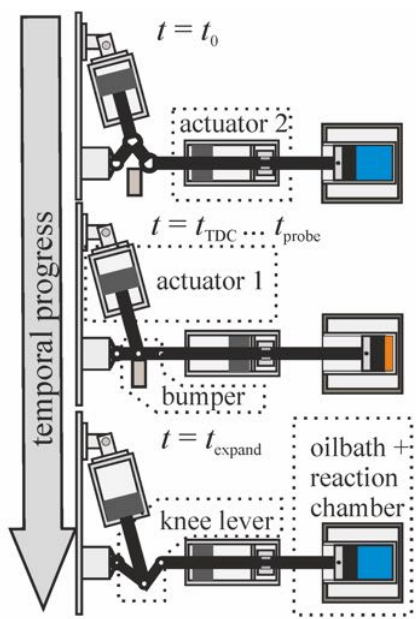

**Figure 1** Schematic construction of the working principle of the RCEM

Figure 1 sketches the RCEM and its working principle. A temperature-controlled oil bath surrounds the piston-cylinder device, providing a well-defined and homogenous temperature distribution. Similar to other RCMs [16-18], the piston has a creviced shape to trap the near-wall gas boundary layer peeled off during the compression. This helps to achieve a more homogenous post-compression temperature distribution within the reaction chamber. The time-resolved in-cylinder pressure is determined by a combination of an absolute pressure gauge (MKS Baratron 121A) and a quartz pressure transducer (Kistler 6061B). The Baratron is used to determine the pre-compression pressure, while the pressure transducer records the whole transient relative pressure history during and after compression; the absolute pressure history is then obtained by calibrating the transducer output with the absolute pressure.

A type K thermocouple is used to measure the pre-compression temperature of the mixture in the reaction chamber. To allow the detection of chemiluminescence, the cylinder head is equipped with an optical access and an optical quartz glass fiber. A photomultiplier (Hamamatsu H10722-210) along with a dichroic short pass filter allows detecting time resolved emissions at selected wavelengths. To record the transient piston position within the cylinder, a potentiometric position sensor (Burster type 7812) is connected to the driving rod. The corresponding geometrical volume history is then derived from the piston position.

A system consisting of pneumatic actuators connected to a knee lever is used to drive the piston (cf. Figure 1). Two pneumatic actuators are connected to the driving rod, one horizontally and another one almost vertically on the hinge of the knee-lever. The use of buffer air tanks keeps the force of the actuators during compression nearly constant. When the actuators are fully pressurized, a pneumatic clamp prevents the driving rod from moving. By releasing the clamp, the experiment and the data acquisition starts.

For compression, the vertical actuator is employed, optionally supported by the horizontal actuator for investigations at high pressures. The knee-lever hits the bumper in its elongated position, thereby locking the piston. Afterwards, the bumper can optionally be removed at any desired time using another pneumatic actuator (not shown in Figure 1) connected to the bumper. The still pressurized vertical actuator pushes the knee-lever down,,thus quickly expanding the gas in the reaction chamber by pulling the piston outwards (Figure 1).

*Defining investigated conditions*

To assign a temperature and pressure to a measured IDT, the corresponding values at TDC are used [4, 11, 16]. As is common practice in RCM experiments [4, 11, 16], the measured pressure at TDC is used together with the initial values of pressure, temperature and mixture composition to calculate the temperature using an adiabatic core (AC) assumption [4, 16]. To achieve a certain desired temperature at TDC, the initial temperature, the compression ratio and the composition of the inert gas in the test mixture can be varied.

*Species profile measurement*

The expansion in the RCEM quickly lowers the in-cylinder gas pressure and temperature, thereby freezing many of the on-going post-compression chemical reactions. This allows a subsequent, convenient ex-situ gas analysis, without requiring high-speed measurement devices. Note that the composition may change during expansion. However, this can be accounted for in numerical simulations, as described in the result part.

To measure species time profiles, several RCEM compression/expansion cycles with identical initial conditions and compression processes are performed. The duration between compression and expansion is varied between different experiments..

By delivering the sample to an optically accessible tube, the absorption spectra of the test gas can be analysed to detect $CH_2O$. The measurement system features a deuterium light source (Ocean Optics D2000, supplying continuous radiation in the region from 200 to 450 nm), and a UV-VIS spectrograph (Ocean Optics USB4000). A more detailed analysis of the test gas is possible with a micro gas chromatograph (Agilent 490 Micro GC). The micro GC is equipped with 3 chromatography columns, namely MS5A, PPU and a 5CB. Among the detectable species are: $H_2$, $O_2$, $N_2$, $CH_4$, $O_2$, $CO_2$, $C_2H_2$, $C_2H_4$, $C_2H_6$, $CH_3OH$, $C_2H_5OH$, $C_3H_6O$, $CH_3OCH_3$.

## 3. Simulation concept

Various numerical models are commonly used to describe RCM measurements and to compare them to simulations [18-21]. A common approach for modelling ignition processes in rapid compression machines is the one described by Mittal et al. [16]. The temporal volume of the AC in the RCM is calculated from a pressure trace of a "non-reactive" measurement. This volume trace accounts for the compression phase and the post-compression heat loss. It is given as an input constraint for the homogeneous reactor model.

In order to improve the description of heat losses, a multi-zone homogeneous reactor model was developed for the RCEM based on the idea of the AC model [16]. For this, the reaction chamber is sub-divided into multiple zones, which are arranged in an onionskin-like fashion (Figure 2). All scalar fields within a zone are assumed to be spatially uniform. The zones have the same instantaneous pressure and they are closed for mass flows. Adjacent zones can exchange heat and each zone can do work by changing its volume. The heat exchanged between two zones is calculated from their temporally varying contact surface, a heat transfer coefficient and the temperature difference between the zones.

Two different heat transfer coefficients are considered here, namely one for inter-zone heat transfer and one for heat transfer between outer zone and wall. The outermost zone also represents the crevice volume and therefore, it has a different heat transfer and surface to volume ratio. To take this into account, the heat transfer coefficient between the wall and the outer zone is allowed to be higher than the one of the inner zones. As a constraint, the sum of all zone volumes has to match the known total volume of the combustion chamber. Then, two values of the heat transfer coefficients are determined numerically, so that the pressure profile of the multi-zone model fits the "non-reactive" pressure measurement best. Consequently, the multi-zone model includes the compression phase. This is necessary, since the compression takes place within a finite time, where chemical reactions can occur [2, 16, 20].

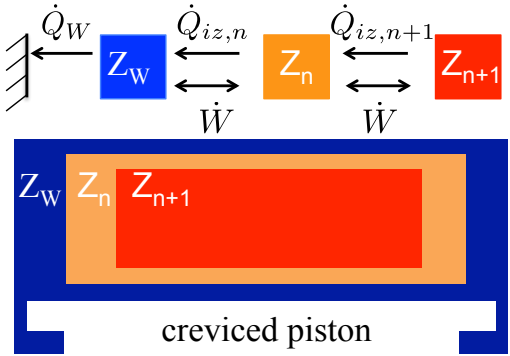

**Figure 2** Structure and working principle of the multi-zone model.

Using the geometrical volume profile as a constraint enables the multi-zone model to also describe an expansion mechanism. This allows the modelling of a quenching process of chemical reactions in the same way as it can be done experimentally with the rapid compression/expansion mechanism.

**3.1 Solution method**

For RCEM model simulations, a homogenous reactor code described by Maas et al. [22] was modified to accommodate the multi-zone model. Here, the zones' pressures are coupled (pressure is the same in all zones), as well as their volumes (the zones' volumes sum up to the instantaneous cylinder volume). The numerical solution of the ODE system resulting from the multi-zone system requires numerous numerical solutions of a large linear equation system. When formulated in the "standard" set of variables (p, V, T), the inter-zone coupling causes the Jacobian of the equation system to not have a banded structure. This makes the numerical solution slower than necessary, by preventing the use of highly efficient numerical algorithms for banded systems. A simple strategy, however, transforms the system to a banded equation system, effecting a speed-up of the numerical solution. The system is transformed by introducing the cumulative sum $(p \cdot V)^*$ of $p \cdot V$ over the zones as a variable, as seen in the formulae below. The resulting equation system is then given, with index $l$ denoting the zones ($l = 1,..,n_z$, with $n_z$ the total number of zones) and index $i$ ($i=1,..,n_s$, with the number of chemical species $n_s$) identifying the chemical species, by:

Species conservation

$$\frac{\partial Y_{i,l}}{\partial t} = \frac{M_i \omega_{i,l}}{\rho_l},$$

energy conservation

$$\frac{\partial u_l}{\partial t} = \frac{p}{\rho_l^2} \frac{\partial \rho_l}{\partial t} + \dot{q}_l,$$

thermal equation of state

$$p_l = \frac{\rho}{\bar{M}_l} RT_l \quad ,$$

auxiliary variable for zone $l=1\ldots n_z$

$$(pV)_l^* = \sum_{i=1}^{l} \frac{m_i RT_i}{\bar{M}_i} \quad ,$$

pressure in zone l

$$p_l = p_{l+1} \quad \text{for} \quad l = 1,\ldots,n_z - 1$$

$$p_{n_z} = \frac{(pV)_{n_z}^*}{V_{\text{tot}}} \quad .$$

The $Y_{i,l}$ denote mass fractions, $M_i$ the molar mass, $\omega_{i,l}$ the molar chemical production term, $\rho_l$ the density, $u_l$ the specific internal energy, $c_{v,l}$ the mixture-averaged specific heat capacity in zone $l$, $p_l$ the pressure, $R$ the universal gas constant, and $m_l$ the mass. The heat which a zone exchanges with its adjacent zones (cf. section on the simulation concept) is given as the net specific heat $q_l$ in the energy conservation. $V_{\text{tot}}$ is the known (transient) total cylinder volume, which is used as a temporal constraint to the system. The differential-algebraic equation system can be solved using the extrapolation code LIMEX [23].

### 3.2 Comparison of the thermodynamic models

Figure 3 compares the temperature- and pressure histories from the AC and from transient multi-zone models, with a varying number of zones, for an "unreactive" DME/CH$_4$/air mixture. Both models use an unreactive measured pressure trace to quantify the heat loss. Accordingly, the pressure traces coincidence quite well. However, for the multi-zone model with less than 5 Zones it is not possible to align the pressure as well as with more zones.

Except for the two-zone model, during a certain period the innermost zone's temperature very closely matches the AC temperature. The duration of this period varies with the total number of zones used in the model for up to 10 zones. For more zones this duration remains nearly constant. As a compromise between computational cost and accuracy the following simulations were conducted with 13 zones.

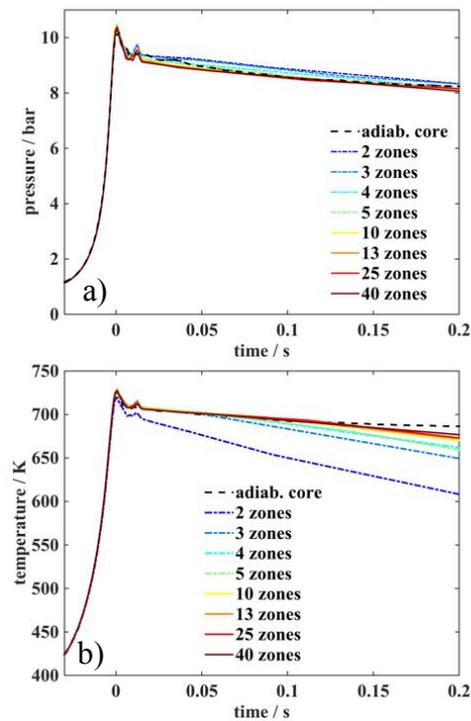

**Figure 3** Temporal pressure and temperature histories of the adiabatic core (dashed line) and the innermost zones of the multi-zone models (solid lines) using different spatial discretization.

To show the effect of including the thermal boundary layer in the simulation, a comparison between measurement, AC- and the multi-zone model is shown in Figure 4. Mittal et al. compared the AC model against CFD-simulations and figured out that the model performs well in predicting the first stage IDT [24]. The multi-zone model shows the same first stage IDT as the AC model. Both models predict earlier ignition than the measurement. However, Mittal et al. also figured out that there is a quantitative discrepancy for predicting the pressure rise in the first stage ignition, thus also for the total IDT using the AC model [24].

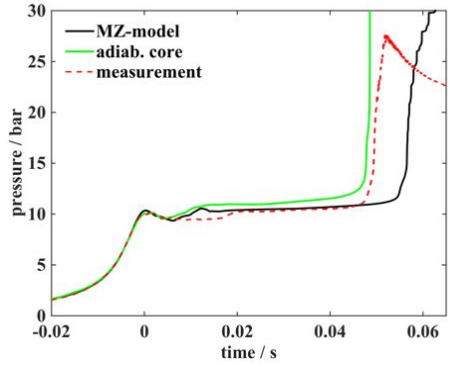

**Figure 4** Temporal pressure histories for a DME/CH$_4$/air mixture with an equivalence ratio of $\phi$=2 of the measurement and the two different numerical models.

In contrast to the AC model [16], the temporal volume change of the particular zones is not predefined. Furthermore, the multi-zone model predicts the occurrence of a near-wall boundary layer and includes the crevice volume, in which no heat release due to first stage ignition takes place. Zones in which heat is released can compress the thermal boundary layer (cf. Figure 5). This leads to a reduced pressure rise, which agrees well with the pressure rise observed in the measurement. In Figure 4, the AC model predicts the total IDT better, since the mechanism was validated against measurements using this model [12].

Beside the description of the time post-first stage ignition, the thermal boundary layer is also important for describing slowly igniting mixtures [20], like for instance the very fuel rich mixtures of this study.

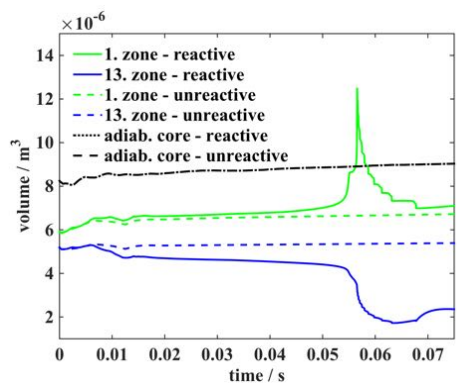

**Figure 5** Temporal volume histories post-compression of the multi-zone and the adiabatic core model for an unreactive and reactive case.

## 4 Results

*Experimental pressure histories during compression/expansion*

Figure 6 shows pressure histories from various compression-expansion sequences, all starting with the same pre-compression conditions. Time $t$=0 s was defined as the end of compression. For each curve, the expansion is initiated at a different time, leading to a rapid pressure decrease. During the compression process, the pressure traces from different experiments match nearly perfectly. For the pressure traces in which expansion took place after the first or main ignition event, also a good agreement of these IDTs is observed. This shows the good shot-to-shot repeatability of the RCEM experiment. As a consequence, ex-situ measurements in gas samples from multiple RCEM experiments with identical initial conditions but different durations between compression and expansion can be used to monitor the evolution of chemical species as a function of reaction time. The samples are averages over the combustion chamber volume.

The slight pressure rise after the expansion (cf. Figure 6) is due to a rebound of the piston, and will be discussed later on.

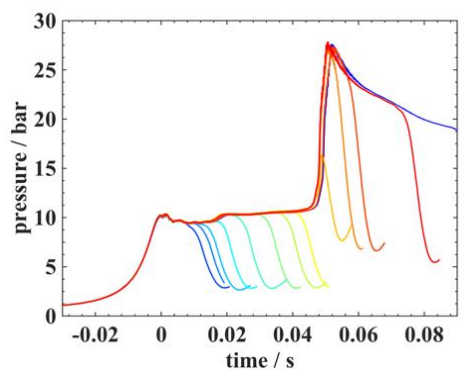

**Figure 6** Pressure traces of 14 RCEM experiments with expansion at different times. Fuel composition: 10 mol% DME 90 mol% $CH_4$; $\Phi$ =2; $T_0$=409 K; $p_0$=0,95 bar; $T_{TDC}$=725,4-728,8 K; $p_{TDC}$=10.07-10.25 bar.

*Effect of expansion on species profiles*

The expansion serves to quickly reduce the pressure and temperature in the mixture, thereby quenching chemical reactions. This quenching is not to be understood as an instantaneous stop of all chemical reactions. In fact, during the expansion, some reactions will be very active, for example recombination reactions of radicals, which have been formed in the post-compression phase. Thus, the whole compression and expansion process has to be modelled.

As mentioned previously, there is a slight surge of pressure after the expansion. This is due to a slight rebound of the piston. This oscillation after the expansion causes a small variation of p, T and V, at a low temperature- and pressure level. This effect is included in the numerical model. Figure 7 compares a pressure history of an exemplary experimental compression-expansion event and the corresponding simulation. It can be seen that the pressure after expansion is predicted well by including the expansion and piston bounce in the model. In addition to the simulated pressure trace, the temporal evolution of some species is shown in Figure 7. It can be seen that by expanding the mixture, the radical $CH_3$ still reacts until it disappears. Stable species retain their values immediately before expansion throughout the expansion. Therefore, the measured post-expansion composition for those species is representative of the situation in the combustion chamber during expansion. However, in the following, the whole post-expansion process including the piston bounce is simulated.

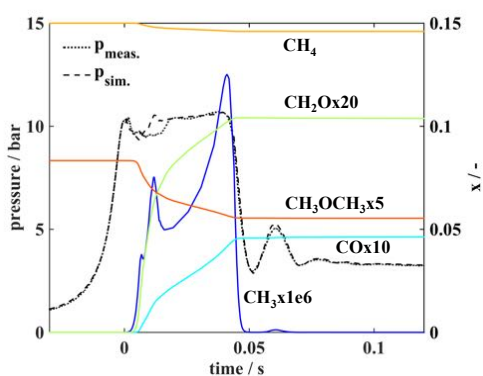

**Figure 7** Comparison between measured and simulated pressure histories of the compression-expansion events. Species time profiles shown over compression-expansion and the post expansion oscillation.

*Comparison of the measured and simulated species evolution*

The temporal consumption of the reactants and the occurrence of intermediate species and products was measured for methane/DME/air mixtures containing 10 mol% DME in fuel, for equivalence ratios of $\phi=2$, 6 and 10. Figure 8 compares measured and simulated consumption of the reactants during the IDT and after ignition. For all three equivalence ratios, DME is consumed too quickly at the beginning. After first stage ignition, DME is consumed more slowly in the simulation. The consumption of oxygen and methane is commonly predicted well by the mechanism. However, for $\phi=6$ a too high consumption of methane is predicted at the main ignition event.

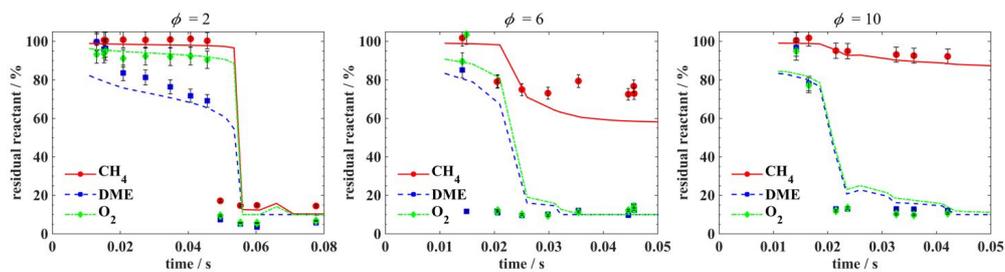

**Figure 8** A comparison of the post compression temporal consumption of the reactants between measurements (symbols) and simulations (lines).

Figure 9 shows the temporal profiles of carbon dioxide, -monoxide, hydrogen and methanol. The amount of carbon monoxide is predicted well over the observation time. For carbon dioxide the simulation predicts little to high values pre-ignition and also post-ignition for the two higher equivalence ratios of $\phi=6$ and 10. Post-ignition the amount of methanol diverge between measurement and simulation, again this is stronger for the higher equivalence ratios. For hydrogen the molar fraction is compared. Both, the pre- and the post-ignition development of hydrogen are predicted good by the mechanism as observed in the measurements. Just the amount of hydrogen post-ignition is slightly overpredicted.

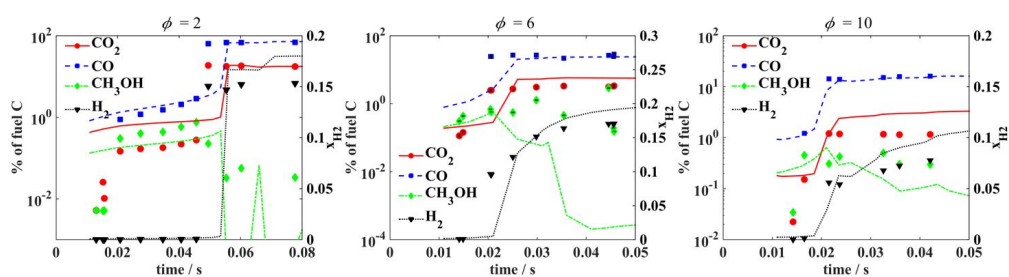

**Figure 9** Comparison of pre- and post-ignition species development between measurements (symbols) and simulations (lines).

Furthermore $C_2H_x$-species are measured and compared to the mechanism in Figure 10. As observed in the measurements and the simulations none of the species is build prior ignition, except a small amount of ethylene. For the two very high equivalence ratios the amounts are described well by the mechanism. The mechanism describes a consumption of the $C_2H_x$-species post-ignition, which isn´t observed in the measurements.

Additionally the amount of formaldehyde, measured via absorption spectroscopy, is shown and compared against the simulation in Figure 10. Prior ignition the amount is slightly overpredicted for $\phi=2$, however the temporal progress of formation and consumption is predicted well. For the higher equivalence ratios no formaldehyde was observed in the measurements prior ignition on the contrary to the simulations. Post-ignition the simulations predict the consumption, whereas in the measurements the amount stayed constant.

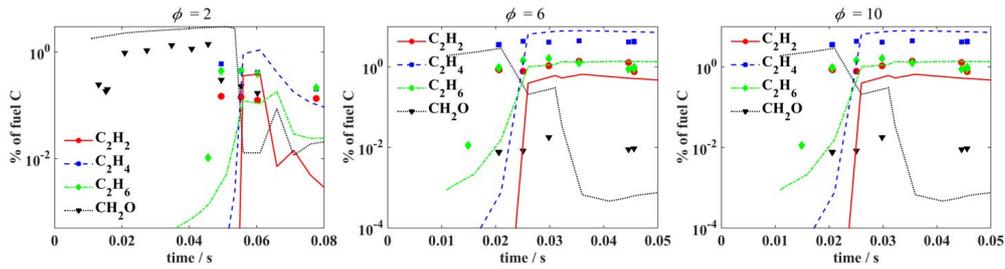

**Figure 10** Comparison of pre- and post-ignition species development between measurements (symbols) and simulations (lines).

## 5. Summary and conclusions

A Rapid Compression Expansion Machine (RCEM) is introduced which extends the usual Rapid Compression Machine (RCM) concept by adding an additional controlled expansion to the process. Comparison of pressure traces recorded in multiple compression/expansion sequences shows the good repeatability of the device. This allows measuring a quasi-temporal species evolution by a series of measurements with expansion at different times relative to the compression.

A numerical multi-zone model was developed to describe the compression and the expansion process of the RCEM was developed. The model is calibrated by non-reactive experiments. Using the model, the thermodynamical processes within the reaction chamber can realistically be described, resulting in better agreement between measured and simulated pressure traces, particularly if two-stage ignition phenomena occur. The multi-zone model is also able to describe the whole expansion process including the pressure oscillation due to the piston bounce.

Measurements using the expansion mechanism for fuel rich DME/methane mixtures were conducted. The results are compared against the multi-zone model. It is shown, that the mechanism is already capable to predict the temporal product and intermediate product formation well. However, by comparing the reactant consumption it could be seen, that DME is firstly consumed too fast and after first stage ignition took place it is consumed slower than observed in the measurements.

**Acknowledgements**

Financial support by the Deutsche Forschungsgemeinschaft within the framework of the DFG research group FOR 1993 "Multi-functional conversion of chemical species and energy" (no. SCHI647/3-1) is gratefully acknowledged.

List of figure captions:

Figure 1 Schematic construction of the working principle of the RCEM.

Figure 2 Structure and working principle of the multi-zone model.

Figure 3 Temporal pressure and temperature histories of the adiabatic core (dashed line) and the innermost zones of the multi-zone models (solid lines) using different spatial discretization.

Figure 4 Temporal pressure histories for a DME/CH$_4$/air mixture with an equivalence ratio of $\phi$=2 of the measurement and the two different numerical models.

Figure 5 Temporal volume histories post-compression of the multi-zone and the adiabatic core model for an unreactive and reactive case.

Figure 6 Pressure traces of 14 RCEM experiments with expansion at different times. Fuel composition: 10 mol% DME 90 mol% CH$_4$; $\Phi$ =2; $T_0$=409 K; $p_0$=0,95 bar; $T_{TDC}$=725,4-728,8 K; $p_{TDC}$=10.07-10.25 bar.

Figure 7 A comparison of the post compression temporal consumption of the reactants between measurements (symbols) and simulations (lines).

Figure 8 A comparison of the post compression temporal consumption of the reagents between measurements and simulations for three equivalence ratios.

Figure 9 Comparison of pre- and post-ignition species development between measurements (symbols) and simulations (lines)

Figure 10 Comparison of pre- and post-ignition species development between measurements (symbols) and simulations (lines)